\documentclass[checkin,showpacs,aps,prb]{revtex4}
\usepackage{makeidx}
\usepackage{listings}
\newcommand{\ba}{\begin{eqnarray}}
\newcommand{\ea}{\end{eqnarray}}

\usepackage{graphicx,color}

\newcommand{\be}{\begin{equation}}
\newcommand{\ee}{\end{equation}}

\newcommand{\et}{{\it et al. }}

\def\prl{{ Phys. Rev. Lett. }}

\newcommand{\clr}{}

\begin{document}




\title{Correlation Between Spin and Orbital Dynamics During
  Laser-Induced Femtosecond Demagnetization}





\author{G. P. Zhang$^*$} \affiliation{Department
  of Physics, Indiana State University, Terre Haute, Indiana 47809,
  USA}

\author{Mingqiang Gu} \affiliation{Department of Physics, Southern
  University of Science and Technology, Shenzhen 518055, China}


\author{Y. H. Bai} \affiliation{Office of Information
  Technology, Indiana State University, Terre Haute, Indiana 47809,
  USA}

\author{T. L. Jenkins} \affiliation{Department
  of Physics, Indiana State University, Terre Haute, IN 47809,
  USA}

\author{Thomas F. George}

\affiliation{Departments of Chemistry \&
  Biochemistry and Physics \& Astronomy, University of
  Missouri-St. Louis, St.  Louis, MO 63121, USA }


\date{\today}

\begin{abstract}
  {Spin and orbital angular momenta are two intrinsic properties of an
    electron and are responsible for the physics of a solid. How the
    spin and orbital evolve with respect to each other on several
    hundred femtoseconds is largely unknown, but it is at the center
    of laser-induced ultrafast demagnetization.  In this paper, we
    introduce a concept of the spin-orbital correlation diagram, where
    spin angular momentum is plotted against orbital angular momentum,
    much like the position-velocity phase diagram in classical
    mechanics.  We use four sets of highly accurate time-resolved
    x-ray magnetic circular dichroism (TR-XMCD) data to construct four
    correlation diagrams for iron and cobalt. To our surprise, a
    pattern emerges.  The trace on the correlation diagram for iron is
    an arc, and at the end of demagnetization, it has a pronounced
    cusp.  The correlation diagram for cobalt is different and appears
    more linear, but with kinks.  We carry out first-principles
    calculations with two different methods: time-dependent density
    functional theory (TDDFT) and time-dependent Liouville density
    functional theory (TDLDFT).  These two methods agree that the
    experimental findings for both Fe and Co are not due to
    experimental errors.  It is the spin-orbit coupling that
    correlates the spin dynamics to the orbital dynamics.
    Microscopically, Fe and Co have different orbital occupations,
    which leads to distinctive correlation diagrams.  We believe that
    this correlation diagram presents a useful tool to better
    understand spin and orbital dynamics on an ultrafast time scale.
    {\clr A brief discussion on the magnetic anisotropy energy is also
      provided.}  }
\end{abstract}

 \maketitle

 \section{Introduction}

The last two decades have seen enormous development in laser-induced
ultrafast demagnetization on the femtosecond time
scale.\cite{eschenlohr2021,ourbook} Beaurepaire \et \cite{eric} in
their pioneering experiment demonstrated that an ultrashort laser
pulse, without an external magnetic field, could demagnetize a
ferromagnet within 1 ps. Under optical excitation, the orbital degree
of the electron is first excited, but most earlier experiments did not
detect orbital
dynamics,\cite{chen2019,siegrist2019,tengdin2018,kampfrath2011,koopmans2010,mathias2012,wietstruk2011}
though theoretically it is long known that both the orbital angular
momentum $L$ (OAM) and spin angular momentum $S$ (SAM) are
important.\cite{prl00,mingsu2015,tows2015,gariglio2019} Accurate
measurements of OAM are recent. The first of such measurement was done
by Bartelt \et \cite{bartelt2007} in Fe/Gd multilayers, where they
employed time-resolved x-ray magnetic circular dichroism (TR-XMCD) to
detect the spin and orbital momenta at Fe's $L_{2,3}$ edge. The
orbital moment reduces earlier.  Boeglin \et \cite{boeglin2010}
employed a circularly-polarized femtosecond x-ray pulse to measure
both spin and orbital angular momenta of Co at its $L_{2,3}$ edge
(between 780 and 800 eV) in a 15-nm Co$_{0.5}$Pd$_{0.5}$ film and
revealed that spin and orbital show different dynamics. The orbital
decreases faster than the spin by 60 fs, and the ratio of
orbital-to-spin changes with time.  Stamm \et \cite{stamm2010} carried
out a difficult measurement of spin and orbital angular momenta of a
thin Ni film and found that OAM and SAM reduce within 130$\pm 40$ fs,
and the spin-orbit interaction increases by $6\pm 2\%$.  Four years
later, Bergeard \et \cite{bergeard2014} employed TR-XMCD to measure
both spin and orbital angular momenta in Co but in
Co$_{0.8}$Gd$_{0.2}$ and Co$_{0.74}$Tb$_{0.26}$ films. They also found
a difference between spin and orbital dynamics, but this does not mean
that spin and orbital dynamics are independent. Instead, they showed
the ratio of orbital to spin has a well-defined trace, within the
experimental error. However, the data in Co$_{0.74}$Tb$_{0.26}$ are
much noisier than those in Co$_{0.8}$Gd$_{0.2}$. Hennecke \et
\cite{hennecke2019} measured OAM and SAM of Fe at the $L_{2,3}$ edge
in a ferrimagnetic Gd$_{25.3}$Fe$_{65.4}$Co$_{9.3}$ film and found
that OAM and SAM decrease equally in magnitude and speed.  However,
their ratio $L/S$ jumps significantly and does not follow a clear
trend, very different from Co.\cite{bergeard2014} Up to now, there is
no consistent understanding of the interesting experimental features.

In this paper, we introduce the simple concept of a spin-orbital
correlation diagram, so even a minute change in OAM and SAM can be
caught on the same graph. As a result, time evolutions of the spin and
orbital leave a single path on the correlation diagram.  We first test
this on two sets of experimental data for Fe in two different
compounds.\cite{bartelt2007,hennecke2019} The results are interesting
and surprising. The spin-orbital correlation diagram produces a simple
picture even when experimental data are noisy. The trace of SAM and of OAM on the diagram is an arc, and
close to the end of demagnetization, both have a cusp. These features
were evidently unknown to those researchers. Then we apply the correlation
diagram to another two sets of data for Co
\cite{boeglin2010,bergeard2014} and find that cobalt's spin-orbital
correlation diagram is very different. It appears more linear and has
kinks. When we try the correlation diagram on Ni,\cite{stamm2010}  the
diagram senses a significant experimental uncertainty in the data
because of its small OAM. To establish our concept, we carry out two
independent first-principles calculations. One is based on
time-dependent density functional theory, and the other is based on
time-dependent Liouville density functional theory which has an
additional self-consistent step. They both show that there is a kink
on the spin-orbital correlation diagram for hcp Co. These kinks are
from the orbital character change during laser excitation, as shown in
the density of states. Then we take another ferromagnetic alloy
FeNi$_3$ as an example. Our calculation indeed reproduces the arc and
cusp for Fe as observed experimentally. These agreements suggest that
it is highly likely that such a correlation between OAM and SAM exists
in nature.  We believe that the spin-orbital correlation diagram is a
useful tool for experimentalists and theoreticians, and presents a new
perspective for ultrafast spin and orbital dynamics for future
research,

The rest of the paper is arranged as follows. In Sec. II, we present
the theoretical formalism on TDDFT and TDLDFT. Section III is devoted
to the results. The first part is devoted to the experimental
results, and the second part is for the theoretical results.  We conclude
the paper in Sec. IV.

\newcommand{\br}{{\bf r},t}
\newcommand{\brt}{{\bf r},t'}
\newcommand{\brp}{{\bf r'},t}

\newcommand{\ik}{i{\bf k}}

\newcommand{\jk}{j{\bf k}}

\newcommand{\lk}{l{\bf k}}

\newcommand{\bk}{{\bf k}}

\section{Theoretical formalism}

\newcommand{\iik}{i,i,{\bf k}}

We employ two different methods to compute the orbital and spin
dynamics, each of which is based on the full-potential augmented plane wave
method. This method is among the most accurate methods.

\subsection{Time-dependent density functional theory}

Since the beginning of time-dependent density functional theory
(TDDFT),\cite{runge1984,burke2005}  it has attracted enormous
attention around the world. The key idea of TDDFT is that one can
simply construct a noninteracting time-dependent single-particle
potential $v(\br)$ that leads to $v$-representable $n(\br)$ as
uniquely determined. There is a similar one-to-one correspondence just
like the ground state between $v(\br)$ and $n(\br)$ for a given
initial many-body state $\Psi_0$.

The fundamental equation is the time-dependent Kohn-Sham equation, \be
i\hbar \frac{\partial \psi_{nk}(\br)}{\partial t} = H \psi_{nk}(\br),
\ee where $\psi_{nk}$ is the wavefunction for band $nk$.  The
Hamiltonian $H$ consists of \be H=\frac{({\bf p}+e{\bf
    A}(t))^2}{2m_e}+v_{ext}(\br) +v_{H}(\br)+v_{xc}(\br) +v_{soc}
(\br) \label{pa}, \ee where the first term is the kinetic energy,
${\bf A}$ is the laser vector potential, $v_{ext}$ is the external
potential interaction, i.e. crystal potential, $v_{H}$ is the Hartree
potential, $v_{xc}$ is the exchange-correlation potential, and
$v_{soc}$ is the spin-orbit coupling.  Here, we use the time-dependent
density functional theory as implemented in the ELK code.\cite{elk}  A
nice feature of this code is that it uses the same method as Wien2k,\cite{wien2k} so we can compare them easily.

In TDDFT, the charge density is computed by \be
n(\br)=\sum_{nk}^{N_{occ}}f_{nk} |\psi_{nk}(\br)|^2, \ee where the
electron occupation $f_{nk}$ is unchanged during time propagation and
fixed at the initial state occupation. In other words, if the state is
initially occupied, it remains occupied to the end of the simulation.
The change in the density is from the wavefunctions. Unoccupied state
participation is indirect. If we include no unoccupied state, we have
no dynamics, even under laser excitation. These unoccupied states
contribute through the dipole or ${\bf p}\cdot {\bf A}$ operator in
equation (\ref{pa}). For this reason, one has to include a number of
unoccupied states in the simulation, even though they do not
contribute to $n(\br)$. The change in density of states is computed
by projecting $\psi_{nk}(\br)$ on to the beginning wavefunction
$\psi_{nk}({\bf r},0)$.  In general, time propagation in TDDFT for a
metallic system is expensive, and the {\tt ELK} code is no
exception. As will be seen below, we have to use an extremely strong
and very short laser pulse.

\subsection{Time-dependent Liouville density functional theory}

An alternative to TDDFT is the time-dependent Liouville density
functional theory (TDLDFT) that we developed.\cite{jpcm16,jpcm17c}
At each time step $t$, we solve the Kohn-Sham equation
self-consistently as the ground state calculation but on a constrained
excited state potential,\cite{jpcm16}  \be \left
[-\frac{\hbar^2\nabla^2}{2m_e}+v_{eff}(\br) \right ]
\psi_{\ik}(\br)=E_{\ik} \psi_{\ik} (\br),
\label{ks}
\ee where $ \psi_{\ik}(\br)$ and $E_{\ik}$ are, respectively, the
eigenstate and eigenenergy of band $i$ and ${\bf k}$ point.  $v_{eff}$
is determined by \be v_{eff}(\br)=v({\bf r})+\int \frac{n(\brp)}{|{\bf
    r}-{\bf r}'|} d{\bf r}'+v_{xc}(\br), \label{pot} \ee where
$v_{xc}(\br)$ is the exchange-correlation potential,
$v_{xc}(\br)=\delta E_{xc}[n]/\delta n(\br)$. The spin-orbit coupling
is included through the second variational principle.\cite{wien2k}
We use the generalized gradient approximation for the
exchange-correlation energy functional.

The density $n({\br})$ is computed from \be
n({\br})=\sum_{\ik}\rho_{\iik}(t)|\psi_{\ik}({\br})|^2, \label{new}
\ee where $\rho_{\iik}$ is the diagonal element of the density matrix
and is also the electron occupation. Our occupation number
$\rho_{\iik}(t)$ is not fixed from the beginning, which is the first
main difference between TDDFT and TDLDFT.  Here $\rho$ is computed by
solving the Liouville equation for a short time step $\Delta
t$,\cite{jpcm16}  \be i\hbar \frac{\partial \rho_{\bk}}{\partial
  t}=[H_0+H_I, \rho_{\bk}], \label{liu} \ee where $H_0$ is the
field-free Hamiltonian, and $H_I=\frac{e}{m_e}{\bf p}\cdot {\bf A}(t)$
describes the interaction between the laser field and system. Here
{\bf p} is the electron momentum operator, and ${\bf A}(t)$ is the
vector potential of laser. After this time step, we plug the new
$\rho_{\iik}$ into equation (\ref{new}) to generate a new charge density
$n({\br})$, so we can generate a new potential through equation (\ref{pot})
for equation (\ref{ks}) to start a new round of self-consistent
calculation. This self-consistent step is another difference between
TDDFT and TDLDFT. In general, the results from both methods are
compatible .

\section{Results and discussions}

\subsection{Concept of spin-orbital correlation diagram}

\begin{figure}
  \includegraphics[angle=0,width=0.9\columnwidth]{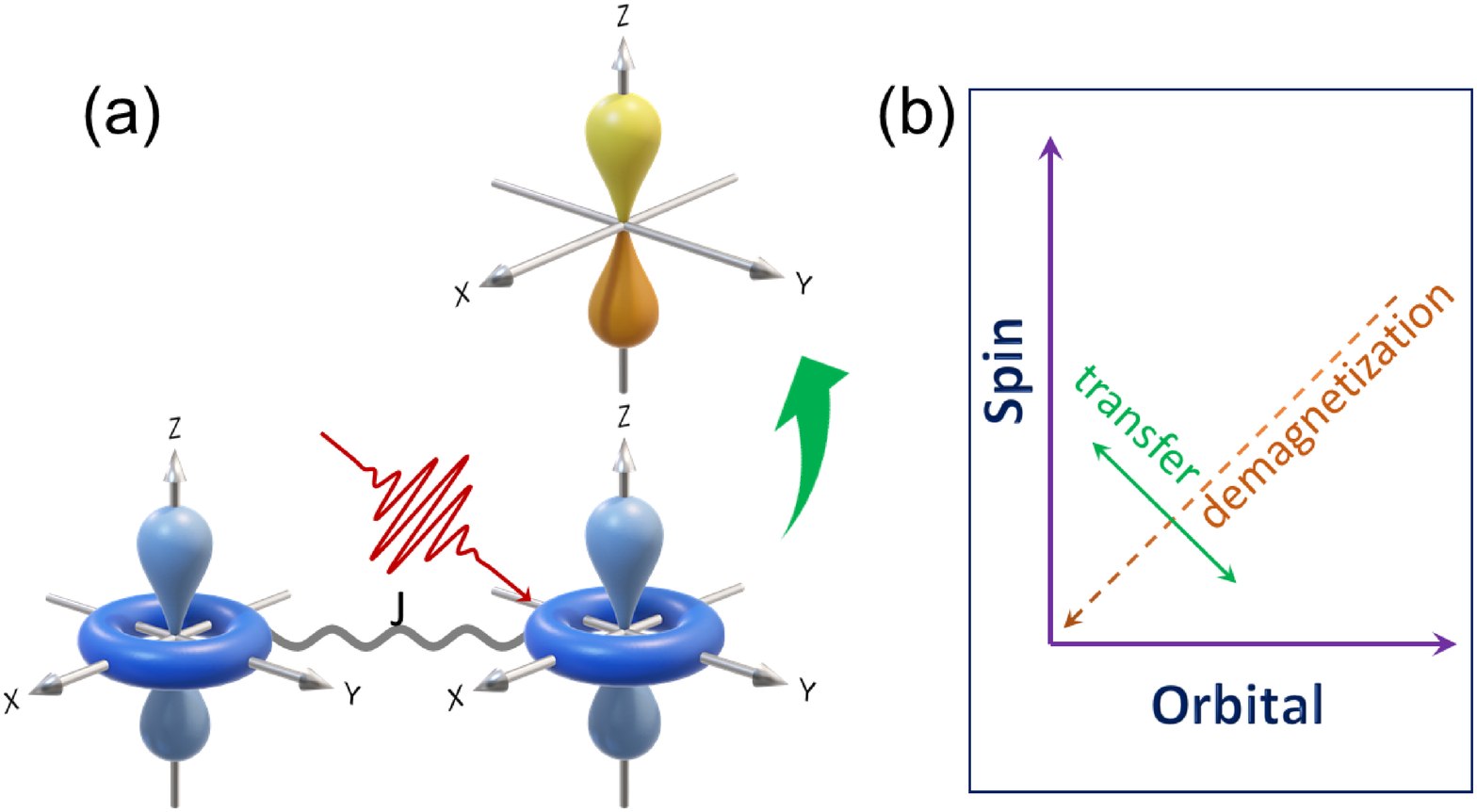}

  \caption{ (a) The laser excites the system first through the orbital
    angular momentum. Exchanged coupled spins are excited through the
    spin-orbit coupling and are further affected by the lattice. The
    orbital dynamics participates in every aspect of ultrafast
    demagnetization. (b) The spin-orbital correlation diagram captures
    both demagnetization (dashed arrow) and angular momentum transfer
    (double arrow), where the time information is factored out.  }
\label{fig0}
  \end{figure}

Optical excitation affects the orbital degree of freedom of an
electron.  Figure \ref{fig0}(a) schematically shows the orbital
angular momentum change during a transition from a $d$ state to a $p$
state. Spins in each orbital are exchange-coupled and are also
affected by lattice vibration. It is this entanglement that requires a
joint study of both spin and orbital dynamics.\cite{oles2004}   Figure
\ref{fig0}(b) schematically outlines the key concept of the
spin-orbital correlation diagram.  For pure demagnetization (see the dashed
arrow), both spin and orbital moments reduce, so their path has a
positive slope. For pure angular momentum exchange (see the double
arrow), the path is transverse, with a negative slope.  The
traditional Land\'{e} $g$-factor is connected to the slope in this
diagram through $M_S/M_O=2/(g-2)$ provided that the orbital moment is
small.\cite{pelzl2003}   In general, these paths are not straight. An
intervention by lattice may distort them. Different experimental
results can now be compared quantitatively, without any ambiguity.

In the following, we start from the existing experimental  data
to build a case that a possible correlation diagram between spin and
orbital dynamics exists. Then we present our first-principles results
to support this conclusion.

\subsection{Experimental evidence}

Decades of experimental research produce a significant amount of data
for spin moment in various materials.  However, most experiments did
not measure orbital moments
explicitly,\cite{chen2019,siegrist2019,tengdin2018,
  kampfrath2011,koopmans2010,mathias2012,wietstruk2011} and those
which measured orbital moments have a large
uncertainty.\cite{stamm2010} The situation changed with the more
recent
investigations.\cite{bartelt2007,boeglin2010,siegrist2019,hennecke2019}
Bergeard \et \cite{bergeard2014} measured both spin and orbital
moments in Co$_{0.8}$Gd$_{0.2}$ accurately, with a much smaller error
bar.

\begin{figure}
\includegraphics[angle=0,width=0.9\columnwidth]{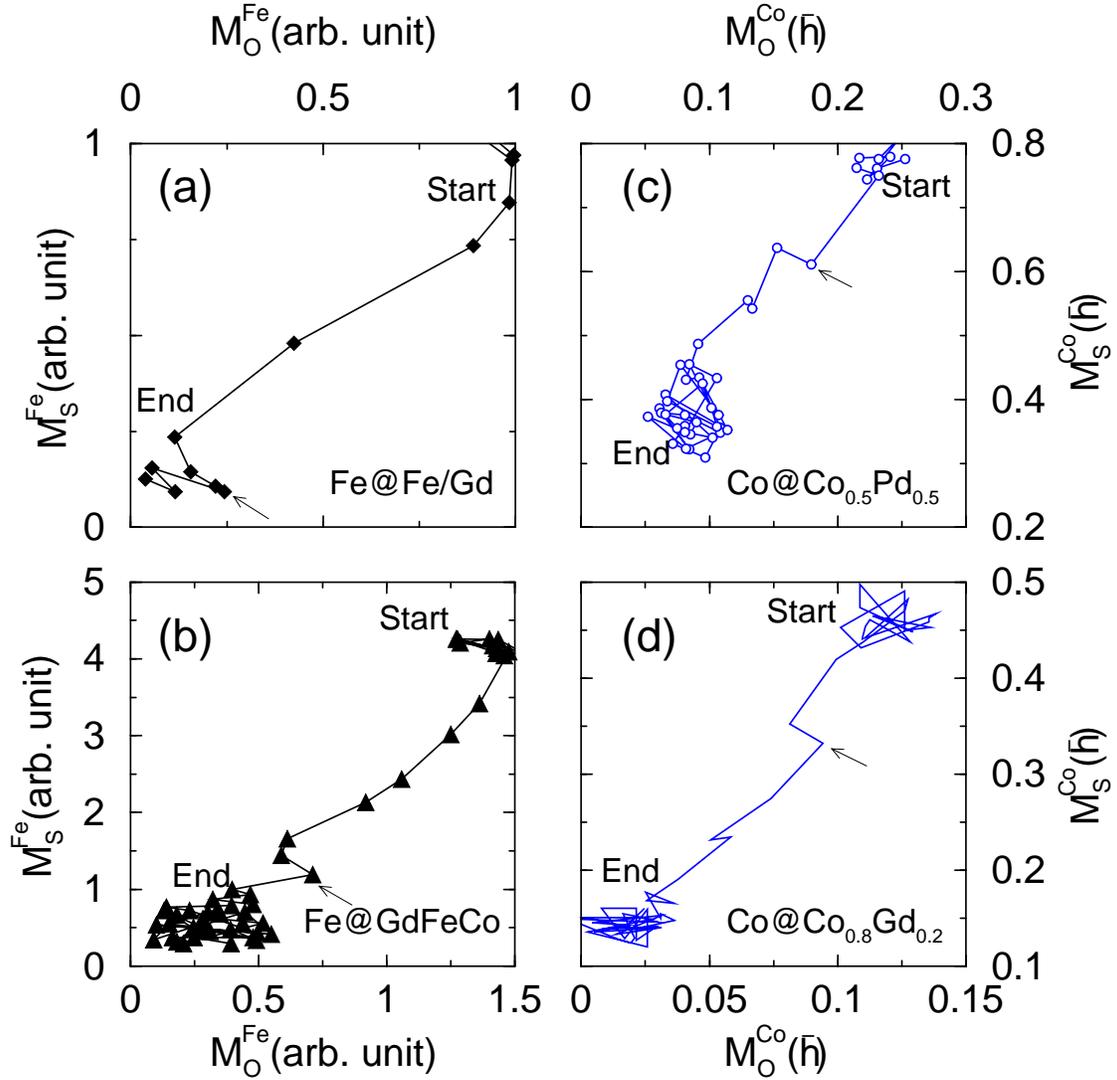}
\caption{ (a) Experimental spin-orbital correlation diagram for Fe in
  Fe/Gd multilayers.\cite{bartelt2007}   The time starts from the
  upper right corner to the lower left corner.  The arrow denotes a
  cusp. {\clr Their experimental time resolution was only 2 ps, but
    their pump had a pulse duration of 60 fs. Therefore we do not
    expect a big effect on our comparison.}  (b) Spin-orbital
  correlation diagram for Fe in GdFeCo.\cite{hennecke2019}   A similar
  cusp is noticed (see the arrow).  (c) Experimental spin-orbital
  correlation diagram for Co in Co$_{0.5}$Pd$_{0.5}$.\cite{boeglin2010}  The arrow highlights a kink. (d) Experimental
  spin-orbital correlation diagram for the Co atom in
  Co$_{0.8}$Gd$_{0.2}$.\cite{bergeard2014}  The arrow highlights a
  kink.}
\label{fig1}
\end{figure}

 Figure \ref{fig1} displays four experimental data of Fe and Co from
 four different materials. All the experimental data points are
 extracted from the original figures using {\sl
   WebPlotDigitizer}.\cite{web} {\sl WebPlotDigitizer} is capable of
 digitizing any data. One only needs to set the $x$ and $y$
 coordinates and the scale, and selects which part of the data needs
 to be extracted.  To demonstrate the accuracy, we plot our extracted
 data against the original data and find excellent agreement.  Figure
 \ref{fig1}(a) plots the spin moment versus orbital moment of the
 experiment done by Bartelt \et \cite{bartelt2007} in Fe/Gd
 multilayers.  {\clr Although their time resolution was 2 ps, their
   pump pulse duration was only 60 fs, which is comparable to our
   theoretical and other experimental durations. In addition, their
   time delay between the pump and probe pulses was only 200 fs, as
   estimated from their data points. For these two reasons, we do not
   expect that their limited time resolution has a big effect on the
   comparison with our theory and other experiments. A major effect
   could be be that their limited time resolution may smear out some
   details of the spin-orbital correlation.}  The starting time is at
 the upper right corner, and the ending time is at the lower left
 corner. This shows that the spin and orbital moments both reduce with
 time. But the spin-orbital trace in the correlation diagram is not
 linear, and in fact it forms an arc, with a cusp at the end of
 demagnetization, as highlighted by an arrow.  This points out that
 there is a phase slip between the spin and orbital dynamics.
 Naturally, all experiments have errors, so we want to see whether
 these features survive in a different material.  Figure \ref{fig1}(b)
 shows the spin-orbital correlation diagram for iron but in
 GdFeCo.\cite{hennecke2019} $\rm Gd_{25.3}Fe_{65.4}Co_{9.3}$ is an
 important material for all-optical spin switching.\cite{stanciu2007}
 Structurally, it is amorphous, quite different from Fe/Gd
 multilayers, yet the spin-orbital correlation diagram is very similar
 to that in Fe/Gd multilayers. One also sees the trace is an arc,
 though it curves upward and has a cusp close to the end of
 demagnetization. Nevertheless, these similarities could be simply
 coincident, so further experimental investigation on iron is
 necessary.

To see whether a similar pattern emerges in other elements, we plot
two additional sets of experimental data but for cobalt. Figure
\ref{fig1}(c) shows the spin-orbital correlation diagram for Co in
$\rm Co_{0.5}Pd_{0.5}$.\cite{boeglin2010}  Different from the above
two experiments which uses the arbitrary units, Boeglin \et provided
the units ($\hbar$) for SAM and OAM, which is very convenient for
theoretical investigations. The spin-orbital trace for Co is
qualitatively different from that for Fe. It appears more linear, but
has kinks (see the arrow in the figure). When we plot the data
from,\cite{bergeard2014}  we notice a similar kink in $\rm
Co_{0.8}Gd_{0.2}$ (see figure \ref{fig1}(d)). The entire spin-orbital
correlation diagram is similar.  The experimental error in the figure
is still slightly too big to make a conclusive statement, but these
qualitative features are interesting. We next examine theoretically
whether this correlation really exists.

\subsection{Physical insights}

Before we present our first-principles result, we would like to have a
basic understanding of why the spin and orbital do not evolve with
time independently.  {\clr First of all, because of spin-orbit
  coupling, {\bf L} is coupled to {\bf S} through $\lambda {\bf
    L}\cdot {\bf S}$, where $\lambda$ is the spin-orbit
  coupling. There is a torque on {\bf L} due to {\bf S}, and vice
  verse.}

We consider a band state $\psi$, which is a linear
combination of orthonormalized $3d$ orbitals
$|d_m\rangle$,\cite{mingsu2015}  \be \psi=\sum_{m=-2}^{m=+2} c_m^\alpha |d_m\rangle
|\alpha \rangle, \ee where $|\alpha \rangle$ is a spin state (spin up
$\uparrow$ and spin down $\downarrow$), and $c_m^{\alpha}$ is the
time-dependent coefficient. Here we assume that the magnetic orbital
quantum number $m$ is still a good quantum number for this example.
The orbital and spin angular momenta are
\begin{widetext}
\ba \langle
\hat{L}_z\rangle &=& \sum_{m=-2}^{m=+2} m \hbar
(|c_m^{\uparrow}|^2+|c_m^{\downarrow}|^2) =\sum_{m=-2}^{m=+2} m \hbar
(|\rho_{m,m}^{\uparrow}|^2+|\rho_{m,m}^{\downarrow}|^2)
 \label{orbital}
 \\ \langle
\hat{S}_z\rangle &=& \sum_{m=-2}^{m=+2} \hbar/2
(|c_m^{\uparrow}|^2-|c_m^{\downarrow}|^2)=\sum_{m=-2}^{m=+2} \hbar/2
\left (|\rho_{m,m}^{\uparrow}|^2-|\rho_{m,m}^{\downarrow}|^2 \right),
 \label{spin}
\ea
\end{widetext}
where $|c_m|^2$ is replaced by diagonal elements of the density matrix
$\rho$ (occupation).  One notices that the spin and orbital angular
momenta share the same density matrix, but whether the spin and
orbital change similarly or not sensitively depends on how
$\rho^{\uparrow}$ and $\rho^{\downarrow}$ change with time.  And this
in turn depends on spin-orbit coupling.  Equation (\ref{orbital})
further reveals that the orbital is prone to the occupation change.
Each occupation $\rho$ at $+m$ has a counterpart $\rho$ at
$-m$. Because they appear simultaneously in the same summation, this
renders the orbital change more radical.  By contrast, the spin term
is insensitive to this rapid change in $\rho_m$ since $m$ does not
appear in the summation explicitly, so the spin change is smoother.
To directly understand the above experimental features, we carry out
numerical calculations.

\subsection{First-principles results}

\subsubsection{TDDFT results}

To confirm the above experimental findings, we use the ELK code
\cite{elk,dewhurst2016} to carry out the time-dependent density functional
calculation.  However, a direct calculation in the time domain with the
same magnetic material as the experimental study is difficult because
of the complexity of the systems. For instance, GdFeCo is
amorphous, which poses a serious challenge. However, from the
above discussion, it is clear that the spin-orbital correlation
diagram is highly element specific, where the same element in different
compounds behaves similarly. We start with hcp Co.

The ELK code uses the full-potential augmented plane wave method, with
the same basis functions as the Wien2k code.\cite{wien2k}  The ELK
code solves the time-dependent Kohn-Sham equation in the real time
domain. Both spin and orbital angular momenta are computed.  Because a
calculation with the same experimental laser parameter is extremely
time consuming, here we employ an ultrashort, ultrastrong laser pulse.
The frequency of the photon energy of our laser pulse is 0.03
a.u. (atomic unit), and the duration is 120 a.u. (atomic unit) or 2.9
fs. The vector potential amplitude is 250 a.u. (atomic unit), which
corresponds to 137 mJ/cm$^2$. 1 a.u. of the vector potential is
$\hbar/(a_0e)$ or $1.24384 \times 10^{-5}$ Vs/m.  To convert the
vector potential amplitude in a.u. to $\rm Vfs/\AA$, we have
$250\times 1.24384$ $\rm Vfs/\AA$, which is much stronger than we use
below ($0.03\rm Vfs/\AA$).  The $k$-mesh is $13\times 13\times 7$,
which is also smaller than the mesh used below ($33\times 33 \times
17$).

Our TDDFT results are shown in figure \ref{fig2}(a). The results start
from the upper right corner and end at the lower left corner. A
similar kink is found in the spin-orbital correlation diagram. This
qualitatively agrees with the experimental findings in
figures \ref{fig1}(c) and (d), but this result alone is not enough to
prove the existence of the spin-orbital correlation.

\subsubsection{TDLDFT results}

\begin{figure}
  \includegraphics[angle=0,width=0.8\columnwidth]{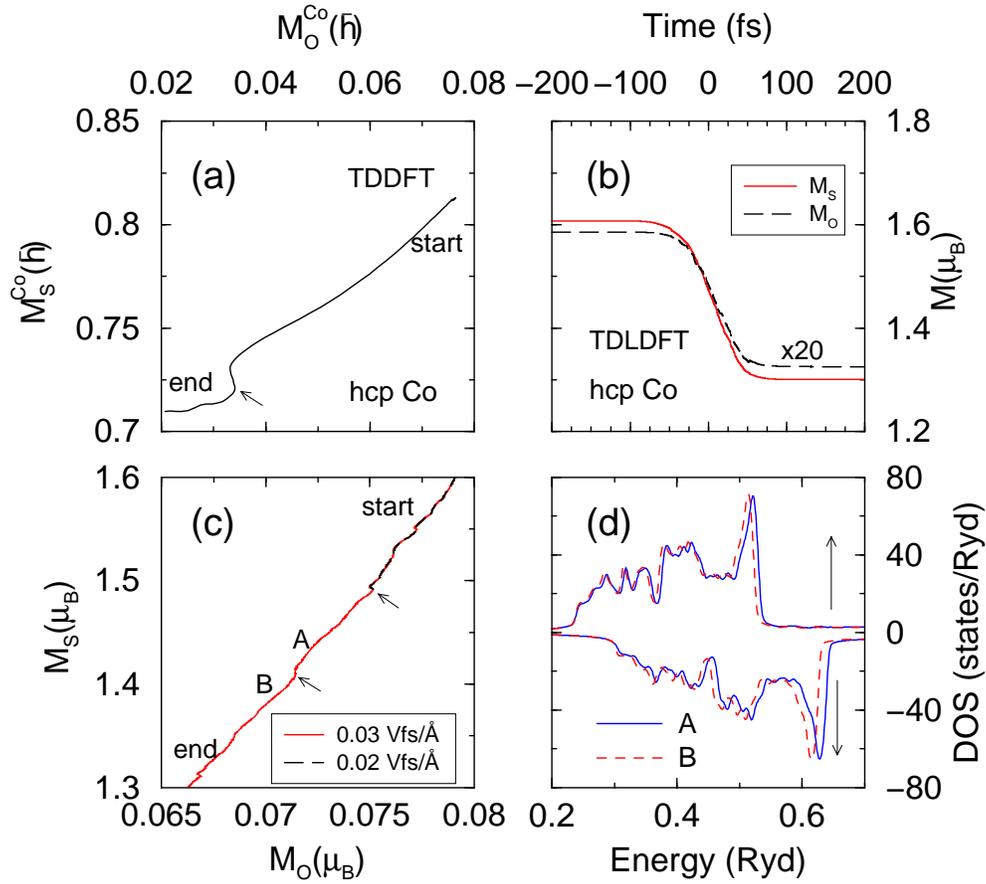}
  \caption{(a) Spin-orbital correlation diagram in hcp Co computed by
    the TDDFT method under an ultrashort, ultrastrong laser pulse.
    (b) Ultrafast spin (solid line) and orbital (dashed line) moment
    reduction upon laser excitation in hcp Co.  The laser field
    amplitude is 0.03 $\rm Vfs/\AA$, the duration is $\tau=60$ fs, and
    the photon energy is $\hbar\omega=1.6$ eV. Here the orbital moment
    is increased by 20 times for an easy view.  (c) Spin-orbital
    correlation diagram under two different laser field amplitudes,
    $A_0=0.03 \rm~ Vfs/\AA$ (solid line) and 0.02 $\rm Vfs/\AA$
    (dashed line).  The arrows highlight the kinks.  (d) Density of
    states for the majority (positive) and minority (negative)
    channels at A and B before and after the kink in (c).  }
\label{exp}
\label{fig2}
  \end{figure}

We employ the time-dependent Liouville density functional theory with
more realistic experimental laser pulses. We use a 60-fs pulse of
$\hbar\omega=1.6$ eV and amplitude of 0.03 $\rm Vfs/\AA$, more than 4
orders of magnitude smaller than our TDDFT pulse.  Such a weak laser
should induce a weaker kink. We plot the spin (solid line) and orbital
(dashed line) moments as a function of time in figure
\ref{fig2}(b). The result for the orbital moment is multiplied by 20 to
have an easy view. One notices that both moments decrease similarly,
but the orbital moment $M_O$ has multiple small steps along the path.
This shows that the orbital moment is much more susceptible to the
orbital character change (see equation (\ref{orbital})).  To amplify the
orbital change, in figure \ref{fig2}(c) we plot the spin-orbital
correlation diagram.  The data (solid line) starts from the upper
right corner.  The small steps in figure \ref{fig2}(b) now produce
distinctive kinks, highlighted by arrows.  A weaker laser field of
0.02 $\rm Vfs/\AA$ leads to a smaller kink (dashed line) that overlaps
with the early portion of the trace of a strong field. These kinks are
indeed smaller.

To understand how the state occupation affects the orbital moment
change in Co as predicted by equation (\ref{orbital}), we choose two points,
A and B, in figure \ref{fig2}(c) before and after the second kink.  We
compute the density of states (DOS) at these two points.  Figure
\ref{fig2}(d) compares the majority DOS (positive axis) with the
minority DOS (negative axis).  We see that the majority and minority
channels both have a significant change.  There is a shift toward the
lower energy at B with respect to A as time evolves. This shift is
larger in the minority channel.  This sudden change alters the orbital
character of the states that contribute to the orbital moment and
subsequently leads to the kink in the spin-orbital correlation
diagram.

\begin{figure}
  \includegraphics[angle=0,width=0.8\columnwidth]{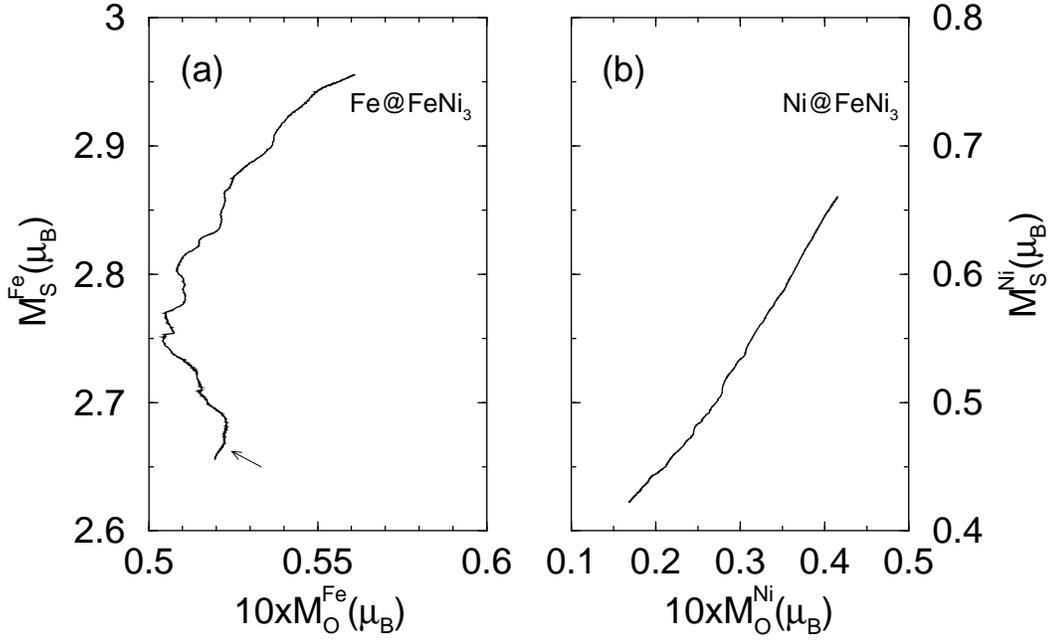}
\caption{Theoretical spin-orbital correlation diagram of (a) Fe and
  (b) Ni in FeNi$_3$. Fe's spin-orbital correlation diagram is an arc
  with a cusp at the end of demagnetization, which agrees with the
  experimental findings in figures \ref{fig1}(a) and (b). Ni's
  correlation diagram is more linear.  The laser field amplitude is
  0.03 $\rm Vfs/\AA$, the duration is 60 fs, and the photon energy is
  1.6 eV.}
\label{fig3}
\end{figure}

We now apply the same concept of the spin-orbital correlation diagram
to FeNi$_3$.  Figure \ref{fig3}(a) shows that the spin-orbital trace
for Fe in FeNi$_3$ is an arc, similar to the experimental findings in
figures \ref{fig1}(a) and \ref{fig1}(b). Interestingly, a cusp is also
found in our theory (figure \ref{fig3}(a)).  In other words, these key
features are found in two experiments \cite{bartelt2007,hennecke2019}
and our theory. This gives us confidence that the spin-orbital
correlation diagram reveals a crucial connection between SAM and OAM.
In figure \ref{fig3}(b), we plot the spin-orbital correlation diagram
for Ni in FeNi$_3$. One notices that the trace is linear. The
experimental result from a prior study \cite{stamm2010} is very noisy
(not shown), probably because the small orbital moment leads to a much
larger error bar.  If we compare our results among Fe, Co and Ni, we
find that when there are more $3d$ orbitals unoccupied, the orbital
change is more radical as seen in Fe. But once they are filled up,
there is not much room for change, so the spin-orbital trace in the
correlation diagram is smoother.  This demonstrates that our proposed
spin-orbital diagram is element specific. We strongly believe that by
looking into various magnetic systems, one may find the spin-orbital
correlation diagram developed here useful for future experimental and
theoretical investigations.  {\clr For instance, the magnetic
  anisotropy energy (MAE) $\Delta E$ is directly related to the
  orbital and spin moments through $H_{\rm soc}=\lambda {\bf
    L}\cdot{\bf S}$,\cite{bruno1989} as already noted
  before.\cite{boeglin2010,stamm2010,bergeard2014} Before laser
  excitation, treating $H_{\rm soc}$ as a perturbation, we have the MAE
  \be \Delta E= \sum_i \frac{|\langle \psi_0|H_{\rm
      soc}|\psi_i\rangle|^2}{E_{0}-E_i}, \label{de} \ee where
  $|\psi_0\rangle$ is the ground state and $\psi_{i}$ is the excited
  state, and $E_0$ and $E_i$ are their respective energies.  And the
  orbital moment is \be \langle {\bf L} \rangle= \sum_i
  \frac{\langle\psi_0|{\bf L}|\psi_i\rangle \langle \psi_i |H_{\rm
      soc} |\psi_0\rangle}{E_0-E_i}. \label{l} \ee Comparing
  eq \ref{de} with eq \ref{l}, one can see the structures of $\Delta
  E$ and $\langle {\bf L}\rangle $ are the same.  If we choose just
  one excited state, $\Delta E/\langle {\bf L}\rangle = \langle
  \psi_0|H_{\rm soc}|\psi_i\rangle / \langle\psi_0|{\bf
    L}|\psi_i\rangle$. States that contribute strongly to $\Delta E$
  or $\langle {\bf L}\rangle$ must have a same quantum number. For
  instance, if $|\psi_0\rangle $ has a $d$ character, then
  $|\psi_i\rangle $ must have a $d$ character, because ${\bf L}$ in
  $H_{\rm soc}$, regardless of its component, does not change the
  orbital angular momentum quantum number $l$ but maybe magnetic one
  $m$ if $L^+$ and $L^-$ are involved. Upon laser excitation,
  $|\psi_i\rangle $ must change its $l$ by 1, say from a $d$ state to
  a $p$ state. This reduces $H_{\rm soc}$, so both MAE and ${\bf L}$
  are reduced as well, which allows spins to deviate from their
  original preferred directions and induces demagnetization across the
  sample.  This is consistent with our recent finding,\cite{mplb21}
  where spins in excited states tend to be disoriented. }

\section{Conclusion}

We have introduced the concept of the spin-orbital correlation diagram
for ultrafast spin and orbital dynamics, where spin angular momentum
is plotted against orbital angular momentum.  As time evolves, the
spin and orbital moments leave a single distinctive trace. First, we
use four different experiment data to show that there is strong
experimental evidence that spin and orbital motions are correlated.
Iron's correlation diagram in both samples is an arc, with a distinct
cusp. Each of these two features is reproduced in our time-dependent
first-principles calculations. The experimental Co correlation diagram
is different. The spin-orbital trace is more linear, but with a
distinctive kink. This is again reproduced by our theoretical
calculation.  Microscopically, we find that the orbital character
change in the density of states in both the spin majority and minority
channels is directly responsible for these features found on the
correlation diagram. As we go from Fe, Co to Ni, the more $3d$
orbitals are occupied, so there is little room for orbital character
change. This explains why in Ni, the trace in its correlation diagram
is highly linear. This demonstrates a high degree of element
specificity.  This correlation diagram provides a new tool to
disentangle complex spin and orbital dynamics for future research.

\acknowledgments

This work was solely supported by the U.S. Department of Energy under
Contract No.  DE-FG02-06ER46304.  Part of the work was done on Indiana
State University's quantum and obsidian clusters.  The research used
resources of the National Energy Research Scientific Computing Center,
which is supported by the Office of Science of the U.S. Department of
Energy under Contract No. DE-AC02-05CH11231.

$^*$guo-ping.zhang@outlook.com.
https://orcid.org/0000-0002-1792-2701

\end{document}